\documentclass[twocolumn,showpacs,preprintnumbers,amsmath,amssymb]{revtex4}
\usepackage{graphicx}% Include figure files
\usepackage{dcolumn}% Align table columns on decimal point
\usepackage{bm}% bold math

\newcommand{\ds}{\displaystyle}
\newcommand{\beq}{\begin{eqnarray}}
\newcommand{\eeq}{\end{eqnarray}}
\newcommand{\beqq}{\begin{eqnarray*}}
\newcommand{\eeqq}{\end{eqnarray*}}

\begin{document}
\preprint{APS/123-PRL}
\title
{Wave function collapse implies divergence of average displacement}

\author{A. Marchewka}
 \email{marchew@post.tau.ac.il}
 \affiliation{Kibbutzim College of Education\\
Ramat-Aviv, 104 Namir Road 69978 Tel-Aviv, Israel}
\author{Z. Schuss}%
 \email{schuss@post.tau.ac.il}
\affiliation{Department of Applied Mathematics\\
Tel-Aviv University, Ramat-Aviv\\
69978 Tel-Aviv, Israel}

\date{\today}

\begin{abstract}
We show that propagating a truncated discontinuous wave function by
Schr\"odinger's equation, as asserted by the collapse axiom, gives
rise to non-existence of the average displacement of the particle on
the line. It also implies that there is no Zeno effect. On the other
hand, if the truncation is done so that the reduced wave function is
continuous, the average coordinate is finite and there is a Zeno
effect. Therefore the collapse axiom of measurement needs to be
revised.
\end{abstract}


\maketitle

\section{\label{sec:intro}Introduction}
We consider the following instantaneous measurement of the
location of a free quantum particle in one dimension
\cite{Feynman}. At a given instant the entire line, excluding a
finite interval, is illuminated to see if the particle is in the
illuminated region. If the result of the measurement is negative,
that is, if the particle is not observed in illuminated region,
the  collapse axiom of quantum mechanics (the collapse or
reduction axiom) asserts that the wave function after the
measurement is the left limit (in time) of the wave function at
the instant of measurement, truncated outside the interval and
renormalized. Alternatively, if the illuminated region is a finite
interval and the measurement establishes that the particle is in
the illuminated interval, without giving its exact location, the
same axiom makes the same assertion as above \cite{CT}. In this
paper we consider the propagation of the wave function after the
reduction. The initial condition for the propagation is
discontinuous, specifically, it has jump discontinuities at the
endpoints of the interval and vanishes identically in the
illuminated region.

We show that propagating the discontinuous wave function by
Schr\"odinger's equation gives rise to non-existence of the
average displacement of the particle on the line or half the line.
One consequence of this phenomenon is that the speed of the
average particle displacement on a given interval is arbitrarily
large if the interval is sufficiently large. In particular, it can
exceed the speed of light. Therefore the collapse axiom of
measurement needs to be revised. On the other hand, if the
truncation is done so that the
reduced wave function is continuous, the average coordinate is finite.\\

\section{\label{sec:2}An exactly solvable example}

To fix the ideas, we consider first the propagation of the initial
rectangle
 \beqq
 \psi_0(x)=\left\{\begin{array}{ll}1&\mbox{if}\quad0<x<1\\&\\0&\mbox{otherwise,}\end{array}\right.
 \eeqq
which is typical of the propagation of more general initial
conditions with jump discontinuities. The propagated wave function
is given by
 \begin{equation}
\Psi (y,t)=\sqrt{\frac m{2\pi i\hbar t}}\int_0^1 \exp \left\{
\frac{im(x-y)^2}{2\hbar t}\right\} \,dx, \label{psi0}
\end{equation}
so its decay for $y\to \infty$ is determined by the asymptotic
behavior of the function erfc($y$) in the complex plane. It is
given by (see \cite{Stegun})
 \beqq
&&\Psi (y,t)=\frac{1}{\pi}\sqrt{\frac{2\hbar
t}{im}}\\
&&\times\left[\frac{\exp\left\{-\ds\frac{imy^2}{2\bar h
t}\right\}}{y}-\frac{\exp\left\{-\ds\frac{im(y-1)^2}{2\bar h
t}\right\}}{y-1}\right]+O\left(\frac{1}{y^3}\right)
 \eeqq
as $y\to\infty$. It follows that
 \beq
 \left|\Psi
(y,t)\right|^2=O\left(\frac{1}{y^2}\right)\quad\mbox{as}\quad
y\to\infty,\label{1y}
 \eeq
hence
 \beq
 \int_0^\infty y\left|\Psi (y,t)\right|^2\,dy=\infty.\label{av}
 \eeq
Although the function $\left|\Psi (y,t)\right|^2$ is integrable,
the function $y\left|\Psi (y,t)\right|^2$ is not absolutely
integrable on the line. Therefore the average $\langle
x(t)\rangle$ does not exist.\\

\section{\label{sec:3}Propagation of a polygon}
We assume that the negative measurement is done outside the finite
interval $[-a,0]$ at time $t=0$, so that wave function is truncated
outside the interval. We consider two cases: (i) the truncated wave
function is discontinuous, and (ii) the truncated wave function is
continuous, though its derivative may be discontinuous. To determine
the wave function at a short time $\Delta t>0$, we approximate the
truncated wave function by an inscribed polygon.

To evaluate the propagated wave function, we divide the interval
$[-a,0]$ into $N$ subintervals and approximate the function
$\Psi_0(x)$ in the $j$-th subinterval by the linear function
$a_jx+b_j$. The free propagation of the approximating function in
each subinterval is given by the integral
 \beq
&&\psi _j(y,\Delta t)=\sqrt{\frac m{2\pi i\hbar \Delta
t}}\label{psytdt}\\
&&\int_{-a+j\Delta
x}^{-a+(j+1)\Delta x}(a_jx+b_j)\exp \left\{ \frac{im(x-y)^2}{2\hbar \Delta t}%
\right\} \,dx.\nonumber
 \eeq
This integral is a sum of two integrals of the form
\begin{equation}
I_0^j=b_j\sqrt{\frac m{2\pi i\hbar \Delta t}}\int_A^B\exp \left\{ \frac{%
im(x-y)^2}{2\hbar \Delta t}\right\} \,dx  \label{Ij0}
\end{equation}
and
\begin{equation}
I_1^j=a_j\sqrt{\frac m{2\pi i\hbar \Delta t}}\int_A^Bx\exp \left\{ \frac{%
im(x-y)^2}{2\hbar \Delta t}\right\} \,dx.  \label{Ij1}
\end{equation}
Changing the variable of integration in eq.(\ref{Ij0}) to $x-y=u$
and then $(A-y)z=u$ in the first and $(B-x)z=u$ in the second
integral gives
\begin{eqnarray}
I_0^j &=&b_j\sqrt{\frac m{2\pi i\hbar \Delta t}}\left[
-(A-y)\int_0^1\exp \left\{ \frac{im(A-y)^2z^2}{2\hbar \Delta
t}\right\} \,dz\right.  \nonumber  \\
&&\nonumber\\
&+&\left. (B-y)\int_0^1\exp \left\{ \frac{im(B-y)^2z^2}{2\hbar \Delta t}%
\right\} \,dz\right] . \label{fint}
\end{eqnarray}
Each one of the two integrals can be evaluated by the asymptotic
formula \cite{Bender}
 \beq
&&\int_0^1\exp \{ixt^2\}dt\sim\label{BO1}\\
&&\nonumber\\
&& \frac 12\sqrt{\frac \pi x}e^{i\pi /4}-\frac
i2e^{ix}\sum_{n=0}^\infty (-i)^n\frac{\Gamma \left( n+\ds\frac 12\right) }{%
\Gamma \left( \ds\frac 12\right) x^{n+1}},\quad x>>1.\nonumber
 \eeq
Applying (\ref{BO1}) to the first integral in eq.(\ref{fint}) with
$x=\displaystyle\frac{m(A-y)^2}{2\Delta t\hbar }$ and to the second
integral with $x=\displaystyle\frac{m(B-y)^2}{2\hbar \Delta t}$,
gives
\begin{widetext}
 \beq
&&I_0^j =b_j\sqrt{\frac{i\Delta t\hbar }{2m\pi }}\left[ \frac{\exp
\ds\left\{ \frac{ im(A-y)^2}{2\hbar \Delta t}\right\}
}{A-y}-\frac{\exp\ds \left\{ \ds\frac{im(B-y)^2 }{2\hbar \Delta
t}\right\} }{B-y}\right]+O\left( \frac{\Delta t^{3/2}}{\left(
A-y\right) ^3}\right) +O\left( \frac{ \Delta t^{3/2}}{\left(
B-y\right) ^3}\right) . \label{I0jO}
 \eeq
\end{widetext}
 This expansion is valid for small $\Delta
t$ and $y$ away from $A$ and $B$ (see below).

Next, from eq.(\ref{Ij1}), we have
\begin{eqnarray*}
I_1^j &=&a_j\sqrt{\frac m{2\pi i\hbar \Delta t}}\left[
y\int_A^B\exp \left\{
\frac{im(x-y)^2}{2\hbar \Delta t}\right\} \,dx\right. \\
&&\\
&&+\left. \int_A^B(x-y)\exp \left\{ \frac{im(x-y)^2}{2\hbar \Delta t}%
\right\} \,dx\right] \\
&&\\ &=&I+II.
\end{eqnarray*}

The first integral, $I$, is similar to $I_0^j$ and is given by
 \beq
&&I =a_jy\sqrt{\frac{i\Delta t\hbar }{2m\pi }}\label{I} \\
&&\nonumber\\ &&\times\left[ \frac{\exp \left\{\ds
\frac{im(A-y)^2}{2\hbar \Delta t}\right\} }{A-y}-\frac{\exp
\left\{\ds \frac{
im(B-y)^2}{2\hbar \Delta t}\right\} }{B-y}\right] \nonumber \\
 &+&O\left( \frac{\Delta t^{3/2}}{\left( A-y\right) ^3}\right) +O\left( \frac{
\Delta t^{3/2}}{\left( B-y\right) ^3}\right)  \nonumber
 \eeq
 (see eq.(\ref{I0jO})). In the second integral, $II$, we set $u=x-y$ and
obtain
\begin{eqnarray}
II &=&a_j\sqrt{\frac m{2\pi i\hbar \Delta t}}\int_{A-y}^{B-y}u\exp
\left\{
\frac{imu^2}{2\hbar \Delta t}\right\} \,du  \nonumber \\
&&\mbox{}  \label{II} =-a_j\sqrt{\frac{i\hbar \Delta t}{2\pi m}}\\
&&\left[ \exp \left\{ \frac{im(B-y)^2}{2\hbar \Delta t}\right\}
-\exp \left\{ \frac{im(A-y)^2}{2\hbar \Delta t}\right\} \right] .
\nonumber
\end{eqnarray}
Now, using eqs.(\ref{I}) and (\ref{II}), we obtain
\begin{eqnarray}
&&I_1^j =I+II =a_j\sqrt{\frac{i\hbar\Delta t}{2m\pi}}\label{I1j}\\
&&\left[\frac{A\exp \left\{ \ds\frac{ im(A-y)^2}{2\hbar\Delta
t}\right\}}{A-y}-\frac{B\exp \left\{\ds\frac{ im(B-y)^2}{2\hbar
\Delta t}\right\} }{B-y}\right]\nonumber\\
&&+O\left( \ds\frac{\Delta t^{3/2}}{\left( A-y\right) ^3}\right)
+O\left( \ds\frac{ \Delta t^{3/2}}{\left( B-y\right) ^3}\right) ,
\nonumber
\end{eqnarray}
For $B=0$ and positive $y$, we obtain
 \beqq
&&I_1^j(B=0)=a_j\sqrt{\frac{i\hbar \Delta t}{2m\pi
}}\,\ds\frac{A\exp \left\{\ds \frac{im(A-y)^2}{2\hbar \Delta
t}\right\} }{A-y}\nonumber\\
&&\\
&&+O\left( \ds\frac{\Delta t^{3/2} }{\left( A-y\right)
^3}\right) +O\left( \frac{\Delta t^{3/2}}{y^3}\right) .
 \eeqq

To calculate the propagation of the entire polygon, we choose the
$N$ vertices at the points
\[
x_j=-a+j\Delta x,\quad \Delta x=\frac aN.
\]
Each side of the polygon is a line $y=a_jx+b_j$, with
\begin{eqnarray}
&&a_j=\frac{\psi (x_{j+1},t)-\psi (x_j,t)}{\Delta x} \nonumber\\
 &&b_j=\psi (x_j,t)-
\frac{\psi (x_{j+1},t)-\psi (x_j,t)}{\Delta x}x_j \label{ajbj}.
 \end{eqnarray}
 Note that
\begin{equation} x_N=0,\quad x_{N-1}=-\Delta
x,\quad b_0=a_0a,\quad b_{N-1}=0. \label{bN0}
\end{equation}
The contribution of the interval $[x_j,x_{j+1}]$ to the integral
(\ref {psytdt}) is given by
 \beqq
&&\psi _j(y,\Delta t)=\sqrt{\frac m{2\pi i\hbar \Delta t}}\\
&&\\
&&\int_{-a+j\Delta
x}^{-a+(j+1)\Delta x}(a_jx+b_j)\exp \left\{ \frac{im(x-y)^2}{2\hbar \Delta t}%
\right\} dx \nonumber.
 \eeqq
We write this as
\begin{widetext}
 \beq
&&\psi _j(y,\Delta t)=\sqrt{\frac m{2\pi i\hbar \Delta t}}\left[
a_j\int_{-a+j\Delta x}^0x\exp \left\{ \frac{im(x-y)^2}{2\hbar \Delta t}%
\right\} \,dx\right.\nonumber \\
&&\nonumber\\ &&\left. -a_j\int_{-a+(j+1)\Delta x}^0x\exp \left\{
\frac{im(x-y)^2}{2\hbar \Delta t}\right\} \,dx+b_j\int_{-a+j\Delta
x}^{-a+(j+1)\Delta x}\exp \left\{ \frac{im(x-y)^2}{2\hbar \Delta
t}\right\} \,dx\right]\label{psijy}
 \eeq
\end{widetext}
for all $j<N$ and fixed $y>0$. Using the integrals (\ref{I0jO}) and
(\ref{I1j}) in (\ref{psijy}), we obtain
\begin{widetext}
\begin{eqnarray}
&&\psi _j(y,\Delta t) \sim\sqrt{\frac{i\hbar \Delta t}{2m\pi
}}\label{malec}\\
&&\nonumber\\ &&\times\left\{ a_j\left[\frac{\left(a-j\Delta
x\right)\exp\left\{\displaystyle\frac{im(-a+j\Delta
x-y)^2}{2\hbar\Delta t}\right\}}{-a+j\Delta x-y}-\right.\right.
\left. \frac{\left( a-(j+1)\Delta x\right) \exp \left\{ \displaystyle\frac{%
im(-a+(j+1)\Delta x-y)^2}{2\hbar \Delta t}\right\} }{-a+(j+1)\Delta x-y}%
\right]  \nonumber \\
&&\nonumber\\
 &&+\left. b_j\left[ \frac{\exp \left\{
\displaystyle\frac{im(-a+j\Delta x-y)^2}{2\hbar
\Delta t}\right\} }{-a+j\Delta x-y}-\frac{\exp \left\{ \displaystyle\frac{%
im(-a+(j+1)\Delta x-y)^2}{2\hbar \Delta t}\right\} }{-a+(j+1)\Delta x-y}%
\right] \right\}=\sqrt{\frac{i\hbar \Delta t}{2m\pi }}\left[
a_jS_j+b_jR_j\right] .\nonumber
\end{eqnarray}
\end{widetext}

First, we define $\alpha =\displaystyle\frac m{2\hbar \Delta t}$,
and using the abbreviation $x_j=-a+j\Delta x$, we rewrite
eq.(\ref{malec}) as
\begin{widetext}
\begin{eqnarray*}
\psi _j(y,\Delta t) &\sim&\sqrt{\frac{i\ }{4\alpha \pi }}\left\{
a_j\left[ \frac{x_j\exp \left\{ i\alpha \left( x_j-y\right) ^2\
\right\} }{x_j-y}- \frac{x_{j+1}\exp \left\{ i\alpha \left(
x_{j+1}-y\right) ^2\right\} }{ x_{j+1}-y}\right] \right.\nonumber\\
&&\nonumber\\
&& +\left. b_j\left[ \frac{\exp \left\{ i\alpha \left( x_j-y\right)
^2\right\} }{x_j-y}-\frac{\exp \left\{ i\alpha \left(
x_{j+1}-y\right) ^2\right\} }{x_{j+1}-y}\right] \right\} \nonumber
\end{eqnarray*}
 or
\begin{eqnarray*}
&&\psi _j(y,\Delta t)\sim\sqrt{\frac{i\ }{4\alpha \pi }} \left\{
\left( a_jx_j+b_j\right) \frac{\exp \left\{ i\alpha \left(
x_j-y\right) ^2\ \right\} }{x_j-y}-\left( a_jx_{j+1}+b_j\right)
\frac{\exp \left\{ i\alpha \left( x_{j+1}-y\right) ^2\ \right\}
}{x_{j+1}-y}\right\} . \nonumber
\end{eqnarray*}
The propagated polygon is
\begin{eqnarray}
&&S =\sqrt{\frac{i\ }{4\alpha \pi }}\sum_{j=0}^{N-1}\left\{\left(
a_jx_j+b_j\right) \frac{\exp \left\{ i\alpha \left( x_j-y\right) ^2\
\right\} }{x_j-y}- \left(a_{k-1}x_k+b_{k-1}\right)
\frac{\exp\left\{i\alpha\left(x_k-y\right)^2\right\}}{x_k-y}\right\}+O\left(
\frac{\Delta t^{3/2}}{y^3}\right)\nonumber
\end{eqnarray}
\end{widetext}
for fixed $y<-a$ and $y>0$. Using the identity $\ $%
\begin{equation}
a_jx_j+b_j=a_{j-1}x_j+b_{j-1}  \label{idej}
\end{equation}
(that expresses the continuity of the polygon at the vertices), we
find that the sum is telescopic and reduces to
\begin{widetext}
\begin{eqnarray*}
S&=&\sqrt{\frac{i}{4\alpha\pi}}\left(a_0x_0+b_0\right)\frac{\exp
\left\{i\alpha\left(x_0-y\right)^2\right\}}{x_0-y}- \sqrt{\frac{i}
{4\alpha\pi}}\left(a_{N-1}x_N+b_{N-1}\right) \frac{\exp \left\{
i\alpha \left( x_N-y\right) ^2\ \right\} }{x_N-y}+O\left(
\frac{\Delta t^{3/2}}{y^3}\right) , \nonumber
\end{eqnarray*}
hence
 \beqq
&&\Psi(y,t)=\sqrt{\frac{i}{4\alpha\pi}}\left[ \frac{\Psi
(-a,0)\exp \left\{i\alpha\left(x_0-y\right)^2\right\}}{x_0-y}
-\frac{\Psi (0,0)\exp \left\{ i\alpha \left( x_N-y\right) ^2\
\right\} }{x_N-y}\right]+O\left( \frac{\Delta
t^{3/2}}{y^3}\right).
 \eeqq
 Thus
 \beqq
|\Psi(y,t)|^2&=&\frac{1}{4\alpha\pi}\left\{\frac{|\Psi
(-a,0)|^2}{(x_0-y)^2} +\frac{|\Psi (0,0)|^2}{(x_N-y)^2}\right.
\left.-2\Re\mbox{e}\left[e^{i\alpha[(x_N-y)^2-( x_0-y)^2]}
\frac{\Psi^* (-a,0) \Psi (0,0)}{(x_0-y)(x_N-y)}\right]\right\}
 +O\left( \frac{\Delta t^2}{y^4}\right).\label{alpha}
 \eeqq
 \end{widetext}
That is, if $\Psi (-a,0)\neq0$ or $\Psi (0,0)\neq0$, then
\begin{equation}\label{1/y}
 |\Psi(y,t)|^2=O \left( \frac{\Delta t}{y^2}\right)
\quad\mbox{uniformly for}\quad y<-a-\delta,\
  y>\delta
\end{equation}
for some $\delta>0$.

Thus there is propagation in short (and long) time and the particle
can be found at any point. As in Section 2, $\langle x(\Delta
t)\rangle$ does not exist. In particular, the average on an interval
$[-a,b]$ at time $\Delta t$ can be made arbitrarily large if $b$ is
chosen sufficiently large. This means that the speed of the average
on a sufficiently large interval can exceed the speed of light (see
\cite{Preskill}). Equation (\ref{1/y}) also means that there is
propagation in short time, so there is no Zeno effect for continuous
time measurements, in contrast to the assertion of existing theories
\cite{Shimizu}.

If, however, $\Psi(y,0)$ is continuous, that is,  $\Psi (-a,0)=\Psi
(0,0)=0$, the above analysis gives
\begin{equation}\label{}
  |\Psi(y,t)|^2=O \left( \frac{\Delta t^3}{y^6}\right)\quad\mbox{uniformly for}\quad y<-a-\delta,\
  y>\delta
\end{equation}
 for some $\delta>0$. This means that there is no
propagation in short time and there is the Zeno effect. In this
case $\langle x(t)\rangle$ is finite.

Note that in case the Hamiltonian contains a finite potential
$V(x)$, the short time propagation, given in \cite{Feynman1},
\cite{Schulman}
\begin{widetext}
 \beqq
&&\Psi(y,t+\Delta t)=\\
&&\\
&&\sqrt{\frac m{2\pi i\hbar t}}\int_{-\infty}^\infty \Psi(x,t)\exp
\left\{ \frac{im(x-y)^2}{2\hbar \Delta t}+i\hbar \Delta t V(x)
\right\}\,dx\left(1+O(\Delta t)\right),
 \eeqq
 \end{widetext}
gives the same-short time result as above. That is, the
discontinuity of the wave function leads to an infinite average.
Note also that the average momentum also diverges.

\section{\label{sec:4}Summary and discussion}
When the mean displacement is infinite the sample average of
measurements of single particle measurements of displacement will
not converge as the sample size increases. This situation is
unacceptable in quantum mechanics, which interprets measurements
through moments (e.g., the uncertainty principle). In particular,
the Ehrenfest theorem does not hold in this case.

There seems to be no natural cutoff that allays this situation,
which leads to the conclusion that the  collapse axiom of
measurements leads to unphysical results. On the other hand, as
the above calculations indicate, a measurement axiom that
truncates the pre-measurement wave function to a continuous
post-measurement wave function circumvents the above mentioned
difficulty. The choice of a continuous post-measurement wave
function is an open problem, which will be discussed separately.

The evolution of a particle, initially distributed uniformly in a
finite potential well, was discussed in \cite{Berry}. Its wave
function develops a fractal space-time structure.  In the case of
continuous measurements inside the infinite well one gets the
quantum Zeno dynamics \cite{Schulman1}. According to our result,
there is no Zeno effect if measurements are negative outside an
interval.

Another consequence of the infinite post-measurement average of
the displacement is that the average can move faster than the
speed of light, which is a new violation of special relativity.
This is also a new violation of quantum information theory,
because the proof that information cannot travel faster then light
concerns finite sets of discrete degrees of freedom, such as spins
\cite{Preskill}, \cite{Page}.


\begin{thebibliography}{99}
\bibitem{Feynman}R.P. Feynman, R.B. Leighton, M. Sands, {\em The
Feynman Lectures on Physics}, vol.3, Addison Wesley, Reading, MA
1965.
\bibitem{CT}C. Cohen-Tanoudji, B. Diu, and F. Lalo\"{e}, {\em Quantum
Mechanics}, Wiley, NY, 1977.
\bibitem{Stegun}M. Abramowitz and I. Stegun, {\em Handbook of
Mathematical Functions}, Dover, NY 1964.

\bibitem{Bender}S.M. Bender and S.A. Orszag, {\em Advanced Mathematical
Methods for Scientists and Engineers}, McGraw-Hill, NY 1978.

\bibitem{Preskill} J. Preskill, {\em Lecture Notes},\\
http://www.theory.caltech.edu/$\sim$preskill/ph219/index.html \#
lecture.
\bibitem{Shimizu}K. Koshino, A. Shimizu, ``Quantum Zeno effect
by general measurements", quant-ph/0411145
\bibitem{Feynman1} R.P. Feynman and  A.R. Hibbs, {\em Quantum Mechanics and
Path Integrals}, McGraw-Hill, NY. 1965
\bibitem{Schulman}L.S. Schulman, {\em Techniques and Applications of Path
Integrals}, Wiley, NY.
\bibitem{Berry}M. Berry, ``Quantum fractals in boxes" {\em J. Phys. A}
{\bf29} p.6617 (1996).
\bibitem{Schulman1}L.S. Schulman, ``Zeno dynamics yields ordinary constraints", {\em
Phy. Rev. A} {\bf65}, 012108 (2001).

\bibitem{Page}D.N. Page, ``The Einstein-Podolsky-Rosen reality is completely described by quantum mechanics",
{\em Phy. Letters A} {\bf91}, p.57 (1982).
\end{thebibliography}
\end{document}